# Collective excitation-mediated transport in nanoscale Josephson junctions that exhibit quantum confinement

Zhengyuan Liu, Sebastian Scherb, Werner M.J. van Weerdenburg[#], Daniel Wegner, Nadine Hauptmann, Alexander A. Khajetoorians*

*Institute for Molecules and Materials, Radboud University, 6525 AJ Nijmegen, the Netherlands*

*Corresponding author: a.khajetoorians@science.ru.nl

[#]Current Affiliation: Fachbereich Physik, Freie Universität Berlin, Germany

**Quantum confinement can strongly modify transport through Josephson junctions. Here, we study local tunneling transport through nanoscale Josephson junction stacks in the Coulomb blockade regime, where the metallic layers exhibit strong vertical quantum confinement. We find that quasiparticle transport is strongly enhanced by a collective excitation mode intrinsic to the junction and localized in the isolated metallic overlayer. We quantify both the collective-mode energy and the Coulomb gap and show that both exhibit strong layer-dependent modulation, consistent with the modulation of the underlying quantum well states. We further investigate how the collective excitation responds to various perturbations, including mechanical motion and an applied magnetic field. Our results suggest that this collective mode is sensitive to quasiparticles near the Fermi level and may therefore provide an indirect probe of the superconducting state.**

The dynamics of Josephson junctions (JJ) are governed by the competition between two energy scales: the charging energy ($E_C$), determined by the junction capacitance, and the Josephson energy ($E_J$), primarily dictated by the critical current. In the small-junction limit ($E_C >> E_J$), transport can be drastically altered [1-3]. It was shown that the critical current can significantly drop in this limit [2], making the phase susceptible to thermal and electromagnetic fluctuations [2,4,5]. It was contested if these ultra-small junctions can be described by the conventional RCSJ model [6], and to what extent scaling these junctions can be used in quantum technologies. At the nanoscale, quantum confinement in the superconducting layers can provide additional mechanisms for modifying superconducting transport [7-9]. In quantum dots, for example, the Josephson current can become strongly dependent on the discrete energy levels of the confined system [10].

For the metallic layers in a standard Josephson junction geometry, quantum confinement becomes prominent when the layer thickness of a given metallic layer approaches the Fermi wavelength. In this limit, the superconducting critical temperature can persist, oscillate, or even increase [11]. For Pb(111), quantum well states (QWS) persists up to ~50 atomic layers, modulating interlayer transport and superconductivity through changes in the quasiparticle density at $E_F$ [12]. In the simplest limit, considering the electronic structure near $\Gamma$ can be primarily described by free-electron-like bands in the plane of the surface with additional sub-band indices introduced by the vertical confinement [13]. In the limit that the lateral size of the metallic layer is small, it can be extremely challenging to determine the superconducting properties due to the presence of strong Coulomb blockade [14]. Coulomb blockade masks conventional measures of superconductivity, such as the superconducting gap. It was proposed that superconductivity can be indirectly probed by a superconducting parity effect in the Coulomb limit [15], where tunneling energies for odd quasiparticle numbers are reduced by the superconducting gap energy.

Here, we study atomic-scale tunneling transport through nanoscale JJ stacks in the Coulomb blockade limit, where the metallic layers exhibit strong vertical quantum confinement. The stacks are grown on Si(111) and consist of two nanometer-thick Pb layers (an over- and an underlayer) with varying thickness and correspondingly layer-dependent QWS, separated by a NaCl dielectric bilayer. The Pb overlayer exhibits significant static Coulomb blockade, where the Coulomb gap strength anticorrelates with the local density of states (LDOS) at $E_F$ associated with the vertically confined states (bilayer oscillation). For Pb overlayers with weak or absent spectral intensity due to the vertically confined state at $E_F$, the tunneling conductance at the onset energy correlated with the Coulomb energy is strongly suppressed. In this case, the conductance is dominated by an additional higher-energy inelastic channel, producing a Coulomb multi-gap structure in STS. This collective excitation, including its energy, is correlated with the spectral weight of the QWS at $E_F$ of the given Pb overlayer. Furthermore, the ratio of conductance between the elastic and inelastic channel is strongly modulated by the junction resistance due to the tip-sample separation. We further demonstrate local excitation of vibrational modes of the Pb overlayer, which modulates the energy of the collective excitations. Using a magnetic field to quench the potential superconductivity in the stack reveals a substantial reduction of the collective excitation energy, on the scale consistent with the expected superconducting pairing energy of confined Pb. We consider various possible origins of this excitation, including potential formation of plasmons and/or polarons in the stack [16-19].

Experiments were performed using a home-built ultra-high vacuum low-temperature STM operated at a base temperature of 1.2 K, and equipped with an out-of-plane 9 T magnet [20]. All experiments were performed using a Nb tip prepared by flashing to ~1500 K. STS was performed using a lock-in technique with a modulation voltage ($V_{mod}$) added to the sample bias ($V_S$). Pb-NaCl-Pb stacks were grown on Si(111) (as sketched in Fig. 1a). The Si(111) substrate was first cleaned by repeated flashing to ~1340 K until the 7×7 reconstruction was observed. The Pb(111) underlayer was grown using a two-step procedure: Pb was deposited onto Si(111) held at ~100 K, and subsequently annealed to room temperature. The underlayers were typically thicker than 40 nm, as confirmed by their QWS fingerprints in STS [21]. NaCl was subsequently grown onto the Pb underlayer at room temperature by thermal evaporation. The resulting NaCl patches, identified by their characteristic rectangular structure [22], typically had lateral dimensions of ~50 nm and a thickness of 2 ML (~0.41 nm), as confirmed by the $\pi$ phase shift of the QWS [23] (Figure S1).

The overlayer was deposited onto the NaCl at room temperature, appearing as hexagonal islands in constant-current images, with typical thicknesses of ~5 nm, as shown in Fig. 1b. Overlayers located on NaCl can be identified by two signatures: (a) a lower-conductance background in constant-current imaging due to the insulating NaCl layer, and (b) spectroscopically, QWS distinct from those of the Pb underlayer (Figure S2). Moreover, all overlayers on NaCl exhibit pronounced Coulomb blockade, as shown in Fig. 1c, with effective gap energies ($E_C$) typically an order of magnitude larger than the superconducting gap energy of Pb ($\Delta$ = 1.42 meV). This suppression of quasiparticle conductance near $E_F$ masks any potential superconducting gap, making it difficult to identify if the overlayer is superconducting at the measurement temperature ($T$ = 1.3 K). In contrast, the Pb underlayer exhibits a superconducting gap consistent with previous reports [24] (Figure S3). We note that the overlayer partially in direct contact with the underlayer do not show such characteristic spectroscopic signatures of Coulomb blockade.

Fig. 1c shows a representative Coulomb gap spectrum measured on a 15-ML-thick stack. The spectrum exhibits a multi-gap structure with four prominent peaks, labeled $\alpha$, $\beta$, $\gamma$ and $\delta$. At typical stabilization parameters ($V_S$ = 50 mV, $I_t$ = 165 pA), the $\beta$ peak at $E_C$+2$\varepsilon$ exhibits substantially higher conductance than the $\alpha$ peak at $E_C$. The appearance of the $\beta$ peaks on both empty/filled states spectra indicates an inelastic tunneling process in which an additional energy $\varepsilon$ is supplied to transport quasiparticle current. Such excitations may originate from collective excitations of the overlayer itself, such as inherent phonons or plasmons, or from excitations between the overlayer and the rest of the stack, such as quasiparticles coupled to collective modes of the underlying layers. The typical value of $\varepsilon$ is on the order of a few meV and discussed subsequently. We note that the multi-gap structure, including the gap attributed to the Coulomb gap ($E_C$) could only be resolved with sufficient energy resolution. This is important as the enhanced conductance at the inelastic peaks can lead to an incorrect determination of $E_C$. Upon further characterization, two additional peaks are observed at higher energies ($\gamma$ and $\delta$). The energy separation between $\alpha$ and $\gamma$ is comparable to the expected electron-phonon coupling energy in Pb [25], while the separation between $\alpha$ and $\delta$ is roughly twice $\varepsilon$, suggesting it is likely associated with a harmonic of the same inelastic excitation (Figure S4).

The conductance of the peaks $\alpha$, $\beta$, $\gamma$ and $\delta$ in the multi-gap structure strongly depends on the tip-sample separation (Fig. 1d-e). The spectra also undergo a rigid shift due to changes in the capacitance between the tip and stack. In the subsequent discussion, we focus on the two main channels, $\alpha$ and $\beta$. At largest tip-sample separation, conductance is primarily dominated by the $\beta$ channel. As the tip approaches the stack, the conductance of the $\beta$ channel saturates and eventually drops. Concomitantly, the conductance of the $\alpha$ channel increases monotonically, and becomes dominant for $\Delta z < -0.33$ nm. The contact point is $\Delta z \sim -0.39$ nm, where a typical superconducting gap spectrum with the expected Josephson peak is recovered (Figure S5). This indicates that quasiparticle transport proceeds more efficiently by driving the collective excitation at furthest tip-sample separation, but through the bare elastic Coulomb channel at closest tip-sample separations. This crossover likely arises from the tip-induced electric field and the resulting potential drop across the stack, which affects the local transport through the stack.

In Fig. 1e, we plot the changes in conductance of the $\alpha$, $\beta$ peaks in comparison to the excitation energy $\epsilon$.  In addition to the discussed changes in conductance, we find that the total energy of the collective excitation reduces monotonically, as the tip is approached, lowering by ~18%. Similar types of reduction in energy, namely red-shifting, has been seen in different limits involving plasmons[26-28].

We subsequently studied the multi-gap structure as a function of Pb overlayer thickness. To correlate the gap structure with thickness, we first characterized the QWS spectra. In Fig. 2a, we illustrates the QWS measured in the empty state regime, above the Coulomb gap onset, revealing a clear bilayer oscillation of the QWS, as expected for Pb in this limit [29]. The QWS of the overlayer differ from those of the underlayer (Figure S2), indicating that the NaCl layer sufficiently decouples the two Pb layers to prevent hybridization.

We next show Coulomb gap spectra for typical stacks in Fig. 2b, with colors and labels corresponding to the QWS in Fig. 2a. We observe two distinct thickness-dependent features: (i) $E_C$ exhibits a bilayer oscillation in its value (gap not centered at 0 mV due to the work-function difference between the overlayer and the underlayer), and (ii) the collective excitation appears only for odd-number-overlayer stacks (see Fig. 2c for zoom-in near the onset of conductance). Concerning (i): we consider the relative difference in conductance due solely to the QWS. The electronic states can be approximated as free-electron in-plane bands with a large effective mass and an additional sub-band index arising from vertical quantization [13,30]. Thus, we can approximate the in-plane contribution constant for all overlayer and variations in conductance primarily reflect QWS crossing $E_F$. The absence of any QWS at $E_F$ is expected to reduce the conductance. Because the Coulomb gap obscures the LDOS at $E_F$, we extract the QWS energy and linewidth (Fig. 2a) and extrapolate their spectral weight into the gap to estimate the LDOS at $E_F$ (Supplementary Note 1 and Figure S6). Using this, we observe that $E_C$ is largest for overlayers with the lowest estimated LDOS at $E_F$ (Fig. 2d), demonstrating that vertical quantum confinement strongly influences the conductance through the stack.

In addition to the thickness dependence of $E_C$, the collective excitation energy $\varepsilon$ shows a similar trend (Fig. 2e). The values of $\varepsilon$ vary from 3 meV to 3.4 meV, across 14 odd-number-overlayer stacks. Since no resolvable excitation is observed for odd-number-overlayer stacks within our experimental resolution (Fig. 2c), we therefore label $\varepsilon = 0$. The collective excitation emerges when the relative LDOS at $E_F$ due to the vertical confinement is weakest. This suggests that, when QWS are absent near $E_F$ and direct quasiparticle transport is suppressed, tunneling becomes more efficient through an inelastic channel via- the collective mode.

To probe the nature of the collective excitation, we intentionally drive mechanical motion of the overlayer. At $V_s$ = 5 mV and $I_t$ = 10 nA, the overlayer can be completely delaminated and transferred to the tip (Figure S7). In tunneling conditions where the tip is closer but not strong enough to delaminate the Pb, we observe additional high-energy inelastic excitations. These appear as replicas of the original multi-gap structure separated by an energy Ω (Fig. 3a), resembling molecular vibrational modes [31,32]. We find that the value of Ω is ~27 meV for the overlayer. This is nearly an order of magnitude larger than the collective excitation energy $\varepsilon$ seen in Fig. 1-2. Furthermore, Ω remains constant for all harmonics (Fig. 3b) and tip-sample separations (Fig. 3c). A spring-model analysis relating Ω to the island volume yields a strong correlation, with the excitation energy decreasing as the overlayer volume (and hence its mass) increases (Supplementary Note 2), confirming its vibrational origin. This observation also rules out an origin of these excitations from underlying excitations, such as from the NaCl layer, since these values should be independent of overlayer size. With these vibrational excitations, we can examine how mechanical motion affects the collective mode. We find that the collective excitation energy $\varepsilon_i$ decreases with increasing vibrational harmonic $i$ (Fig. 3e), indicating that mechanical motion of the overlayer, the overall voltage difference, and its quasiparticles strongly modify the energy of the collective excitation. This suggests that the quasiparticles within the overlayer play an important role in this collective excitation.

We observe a strong magnetic field dependence of the collective excitation. In Fig. 4a, we show the multi-gap structure of four odd-number-overlayer stacks measured at $B$ = 0 T and $B$ = 2 T. We note that the upper critical field of overlayer in this size regime is typically far below 2 T [33,34]. Therefore, we assume that the applied magnetic field is sufficient to quench superconductivity in both the over- and underlayer. Moreover, superconductivity has been seen in Pb clusters down to ~8600 atoms [35]. The presence of the collective excitation in a strong out-of-plane field indicates that the nature of the collective excitation is not solely due to the superconducting state itself, such as Carlson-Goldman modes [36]. Nevertheless, the excitation energy is strongly reduced by the magnetic field. The reduction in $\varepsilon$ likely arises from the additional quasiparticle conductance at $E_F$ introduced by the quenching of the superconducting state of the overlayer, similar to the absence of the observation of the collective excitation for even-number-overlayer stacks. We define $\Delta\varepsilon = \varepsilon(0T) - \varepsilon(2T)$ and compare $\Delta\varepsilon$ for overlayers of different sizes (Fig. 4b). Remarkably, $\Delta\varepsilon$ is comparable to the superconducting gap energy of Pb in this volume limit [15,34,35] and decreases for smaller volume. In the case that the collective excitation is sensitive to the quasiparticle conductance at $E_F$, the reduction of $\Delta\varepsilon$ for smaller overlayers may reflect the suppression of superconductivity due to quantum confinement [15,34,35].

When applying different stabilization conditions (*e.g.* higher $V_S$), the tunneling rate can become comparable to the quasiparticle lifetime on the overlayer, bringing the system into the Coulomb peak regime. Here, discrete peaks correspond to the addition of successive quasiparticle events (1*e*, 2*e*, etc.). It has been shown that in the case of an even number of quasiparticles there can be an energy gain equal to the superconducting gap energy, for even parity [15,37]. We measure the Coulomb peaks with and without an applied magnetic field and extract their energy spacings using the bare Coulomb peaks as the onset (Fig. 4c-d). The energy spacings exhibit a clear even–odd oscillation, suggesting the presence of a superconducting parity effect. However, the oscillation amplitude increases rather than decreases under magnetic field. This indicates that the values of these peaks are most likely not sensitive to the superconducting state of the overlayer, as the applied magnetic field strength is enough to quench any present superconducting state. Therefore, in this system, the origin of the apparent oscillation has a different origin.

In conclusion, we observe a collective excitation in the Coulomb blockade limit for nanoscale Josephson junction stacks, where strong quantum confinement governs the electronic structure of the metallic layers. The Coulomb gap is strongly correlated with the confinement-induced LDOS at $E_F$, exhibiting a bilayer oscillation. The transport is dominated by the collective excitation when the LDOS at $E_F$ is suppressed, namely in layers with an odd thickness value. The collective excitation energy depends strongly on the LDOS at $E_F$, and is not observed in even layers. We show that this behavior is also modified by mechanical motion of the overlayer. Applying a magnetic field strongly reduces the excitation energy, likely through the emergence of quasiparticle states at $E_F$, suggesting a possible indirect probe of superconductivity in the isolated island. We propose that the nature of this collective excitation most likely involves (i) collective motion of the quasiparticles within the overlayer, *e.g.* low-energy plasmons that are quasi-2D, or (ii) coupling between quasiparticles in the overlayer and collective excitations in the layer underneath (*e.g.* phonons). Finally, the multi-gap structure complicates the identification of a superconducting parity effect, and we do not observe a significant reduction of the charging energy for various parity under magnetic field.

**Acknowledgements**

This project was supported by the European Research Council (ERC) under the European Union's Horizon 2020 research and innovation program (grant no. 818399 and 947717). This publication is part of the project "What can we 'learn' with atoms?" (with project number VI.C.212.007) of the research program VICI which is (partly) financed by the Dutch Research Council (NWO).

**Author contributions**

ZL and AAK wrote the paper, with all authors providing insight and comments into the manuscript. ZL, SeS, and WMJvW acquired the experimental data. ZL, SeS, WMJvW, and AAK all contributed to the design of the experiments. ZL, SeS, and WMJvW performed the experimental analysis, with all authors contributing to the discussion of the results of the analysis. AAK, DW and NH supervised the experiments. All authors participated in discussions about the scientific interpretation of the results.

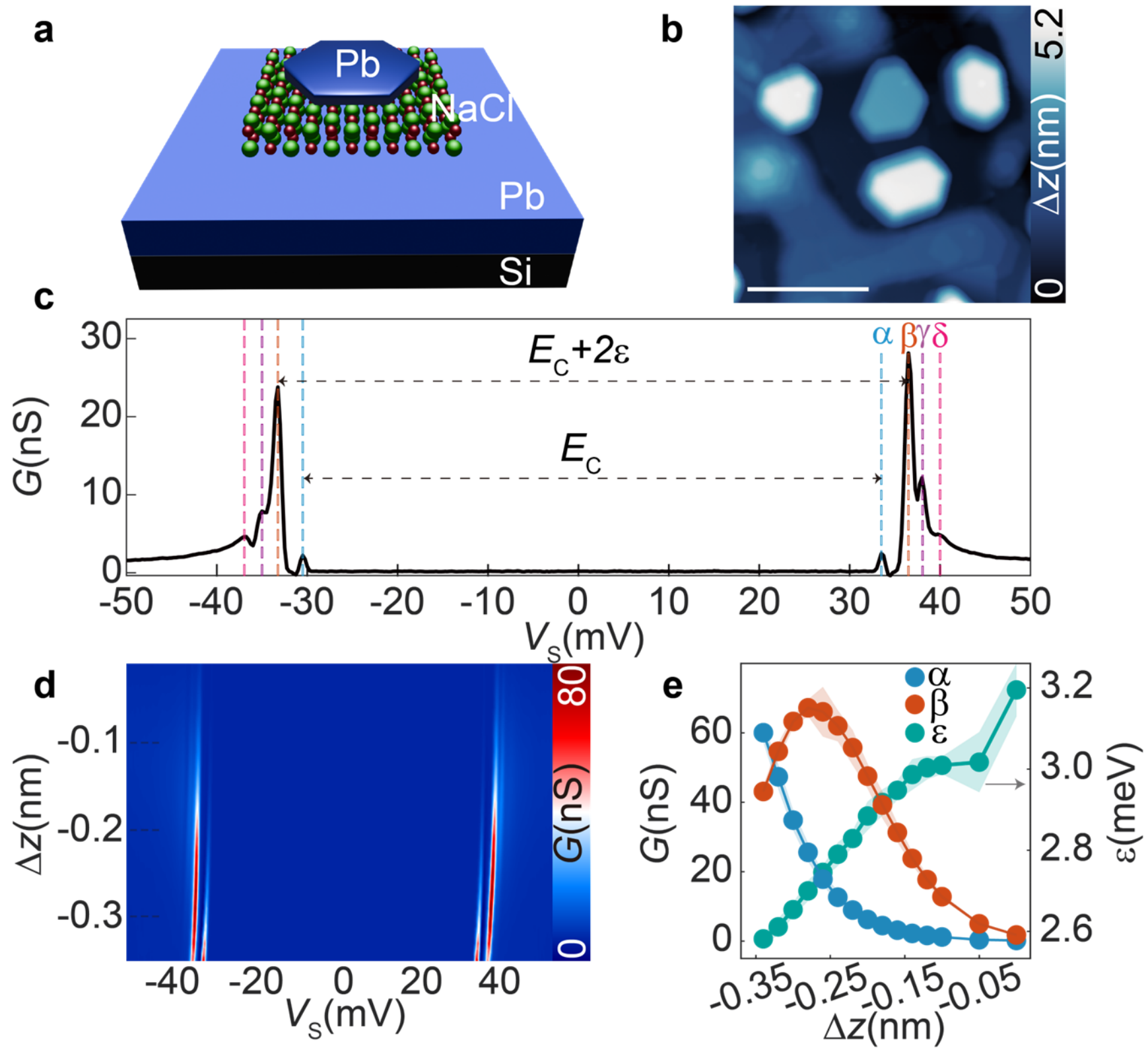

**Figure 1. Coulomb blockade and collective excitations in Pb/NaCl/Pb stack.**
**a** Schematic of the Pb/NaCl/Pb stack grown on Si(111). **b** STM image showing four stacks with overlayer of thickness that ranges from 11 ML to 17 ML ($V_S$ = 2 V, $I_t$ = 20 pA, scale bar is 10 nm). **c** Point d$I$/d$V$ spectra taken on top of a typical stack with a 15 ML overlayer ($V_S$ = 50 mV, $I_t$ = 165 pA, $V_{mod}$ = 500 $\mu$V). **d** Point d$I$/d$V$ spectra taken on the same stack as in **c** at different tip-sample separation (Δz). Δz = 0 refers to stabilization at $V_S$ = 150 mV, $I_t$ = 200 pA. **e** Intensity of peak $\alpha$ and $\beta$ and collective excitation energy $\varepsilon$ as a function of Δz. Error bars in **e** obtained from Gaussian fitting of each peak and indicated by the shaded regions.

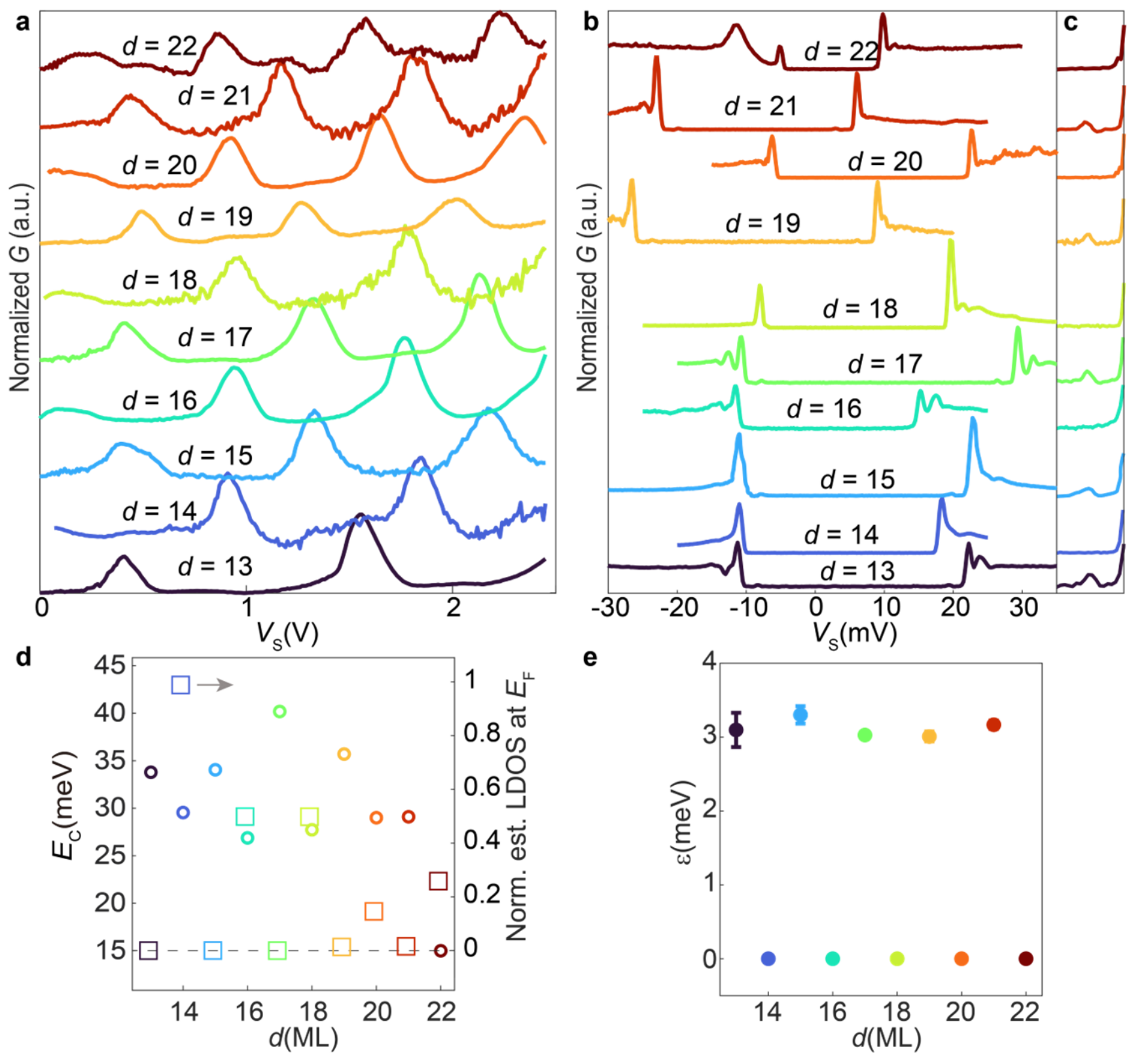

**Figure 2. Quantum oscillations of Coulomb gap and the collective excitation.**

**a** Normalized point d$I$/d$V$ spectra taken on stacks with overlayer thickness ranges from 13 ML to 22 ML ($V_S$ = 2.5 V, $I_t$ = 200 pA, $V_{mod}$ = 10 mV). **b** Normalized point d$I$/d$V$ spectra taken on the same stacks as in **a**. ($V_S$ = 30 mV, $I_t$ = 200 pA, $V_{mod}$ = 500 μV). **c** Zoomed-in spectra for better view of the inelastic excitation peaks. **d** Coulomb gap ($E_C$) and normalized estimated-LDOS at the Fermi energy as a function of overlayer thickness. **e** Collective excitation energy $\varepsilon$ as a function of overlayer thickness.

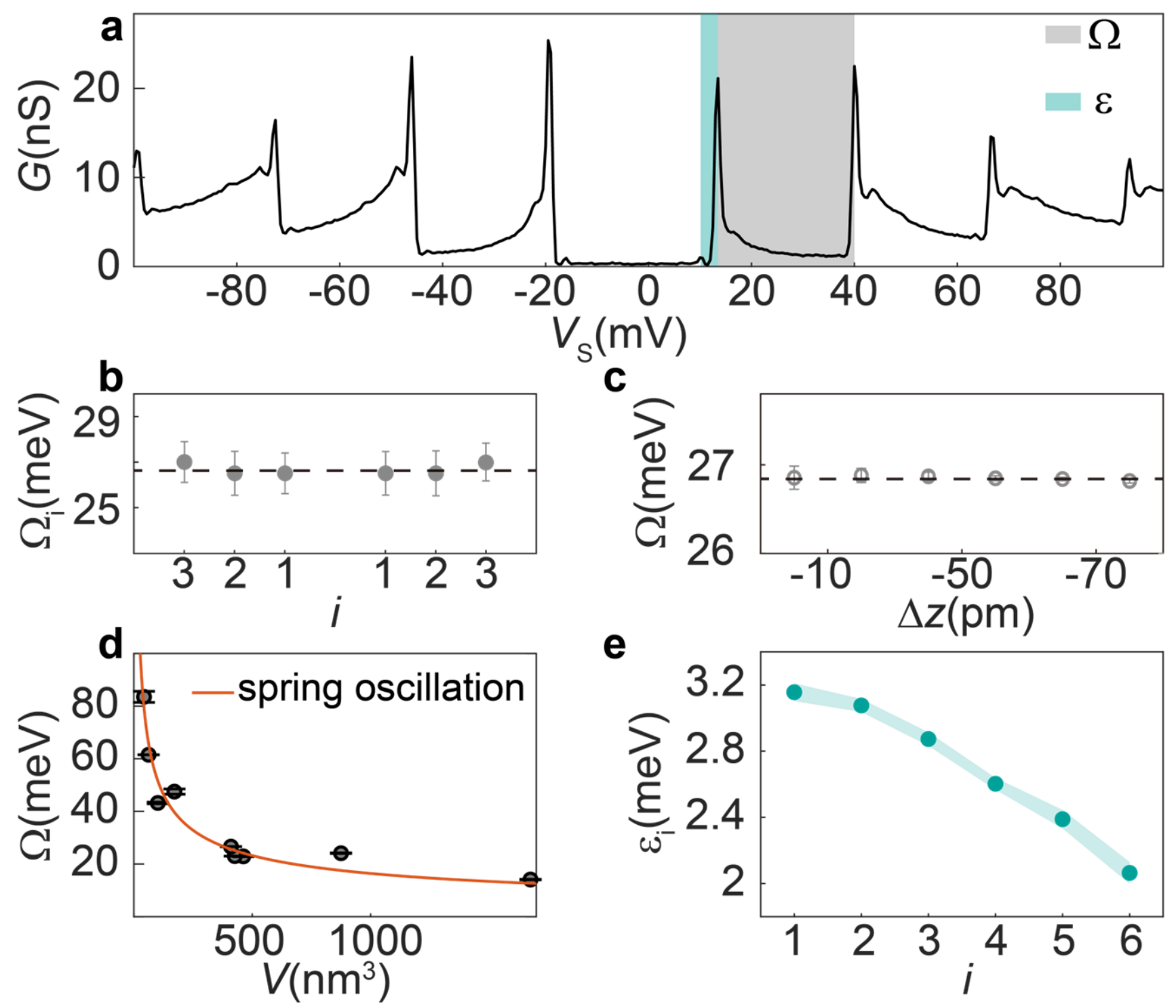

**Figure 3. Mechanical motion of the overlayer and its coupling to the collective excitation.** **a** Typical point d*I*/d*V* spectroscopy taken on top of a stack with a 17 ML overlayer ($V_S$ = 100 mV, $I_t$ = 600 pA, $V_{mod}$ = 200 $\mu$V). Shaded regions indicate the energies for collective excitations and phonon excitations respectively. **b** Phonon energy Ω for different harmonic order. **c** Phonon energy Ω as a function of Δz. **d** Phonon energy Ω as a function of overlayer volume. Red line is a fit based on classical spring oscillation. **e** Collective excitation energy $\varepsilon$ for different harmonic order. Error bars are obtained from the gaussian fitting of each peak and indicated by the shaded regions.

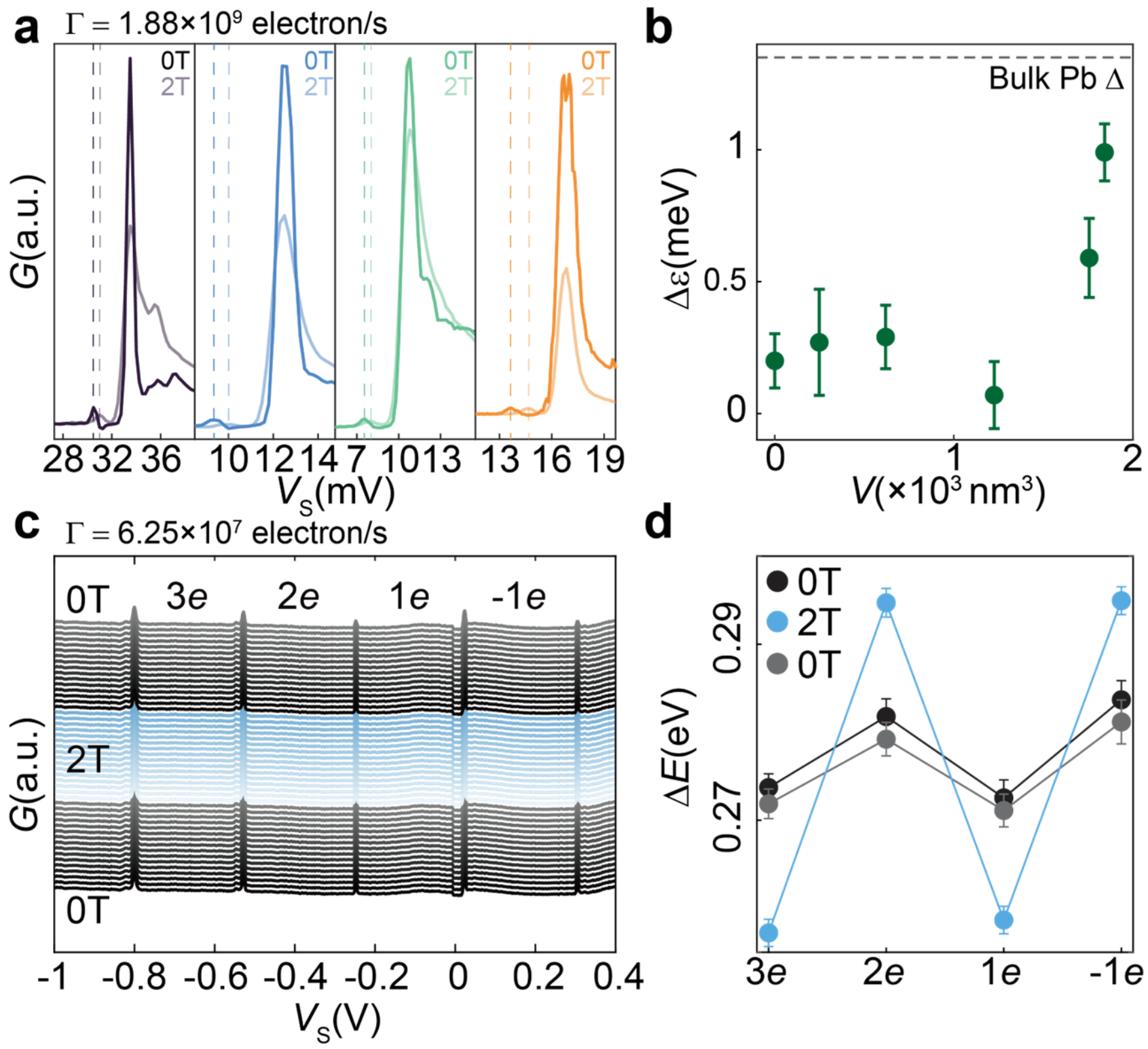

**Figure 4. Quantifying the collective excitation behavior in magnetic field.**
**a** Normalized point d*I*/d*V* spectra taken on four different stacks without field, with out-of-plane magnetic field ($B$ = 2 T, $V_S$ = 30 mV, $I_t$ = 200 pA, $V_{mod}$ = 500 $\mu$V). Dashed lines indicate the energies of peak $\alpha$. **b** Reduction of collective excitation energy $\Delta\varepsilon = \varepsilon(0T) - \varepsilon(2T)$ as a function of overlayer volume. **c** Point d*I*/d*V* spectra taken on top of a typical stack with 17 ML overlayer without field, with out-of-plane magnetic field ($B$ = 2 T), and retaken without field ($V_S$ = 1 V, $I_t$ = 200 pA, $V_{mod}$ = 2 mV). Under each field 16 spectra were taken over a 1 nm$^2$ area to minimize the spatial variations. Excess charge on the overlayer is labeled with ±1*e*, ±2*e*, ±3*e* etc. **d** Coulomb peak spacing as a function of excess charge on the overlayer.

Supplementary Information for

# Collective excitation-mediated transport in nano-scale Josephson junctions that exhibit quantum confinement

Zhengyuan Liu, Sebastian Scherb, Werner M.J. van Weerdenburg#, Daniel Wegner, Nadine Hauptmann, Alexander A. Khajetoorians*

*Institute for Molecules and Materials, Radboud University, 6525 AJ Nijmegen, the Netherlands*

*Corresponding author: a.khajetoorians@science.ru.nl

#Current Affiliation: Fachbereich Physik, Freie Universität Berlin, Germany

# Contents

## Supplementary note 1: Linear extrapolation of the quantum-well states to $E_F$

Due to the strong Coulomb blockade (*i.e.* conductance gap), the observation of the quantum-well states (QWS) could not always be followed down to $E_{F.}$ Therefore, its contribution to the local density of states (LDOS) at $E_F$ was determined by extrapolation.

In the experimentally accessible range closest to $E_F$, the QWS energy exhibits an even spacing between subsequent states, namely:

$$E_{n+1} - E_n = E_n - E_{n-1}$$

where $n$ is the sub-band index. Thus, we get

$$E_0 = 2E_1 - E_2$$

where $E_0$ is the energy for the QWS closest to $E_F$, $E_1$ is the energy for the first QWS above $E_F$ and $E_2$ is the energy for the second QWS above $E_F$. $E_1$ and $E_2$ are determined by fitting the measured QWS peak with Gaussian. The peak height is estimated based on the how the intensity decays and linewidth of the extrapolated QWS is assumed to be the same as the QWS closest to $E_F$.

The fit was performed only using islands that exhibited QWS positions that had significant intensity and could therefore be reliably identified. This procedure provides an estimate of the QWS Fermi-level crossing even when the state becomes too weak or broad to be resolved experimentally in the immediate vicinity of $E_F$.

## ■ Supplementary note 2: Spring oscillation model for Pb islands

To examine whether the high-energy inelastic excitation Ω originates from mechanical motion of the top Pb island, we model the island as a classical harmonic oscillator with an effective spring constant $\kappa_{eff}$ and mass $m$. This is based on the observation that the underlying frequency is correlated with the island mass, and that with sufficient tunneling conditions, the island can be delaminated. The resonance frequency is

$$\omega_0 = \sqrt{\frac{\kappa_{eff}}{m}}$$

corresponding to an excitation energy

$$\Omega = \hbar\omega_0 = \hbar\sqrt{\frac{\kappa_{eff}}{m}}$$

Approximating the mass of the Pb island as $m = \rho V$, where $\rho$ is the mass density of Pb and $V$ is the island volume, gives $\Omega = AV^{-1/2}$, where $A = \hbar\sqrt{\frac{\kappa_{eff}}{\rho}}$ is treated as a fitting parameter. Thus, for an approximately thickness- and size-independent effective restoring force, the classical oscillator model predicts that the excitation energy decreases as $V^{-1/2}$.

The island volume was estimated from its lateral area and apparent height, correlated with the QWS spectra. Potential strain is neglected. The determined excitation energies were then fitted to $\Omega = AV^{-1/2}$. The observed decrease of Ω with increasing island volume is consistent with the expected mass dependence of a mechanical mode. In addition, the approximately constant energy spacing between successive replicas of the multi-gap structure is consistent with harmonics of the same vibrational excitation. These observations support the assignment of the high-energy inelastic features to mechanically driven motion of the top Pb island.

We emphasize that this model provides a phenomenological description rather than a microscopic treatment of the island dynamics. The effective spring constant incorporates the restoring forces associated with the Pb–NaCl interface, tip–island interaction, and possible deformation of the island. Nevertheless, the characteristic $\Omega \propto V^{-1/2}$ dependence provides a simple test of the mechanical origin of the observed excitation.

■ **Supplementary figures**

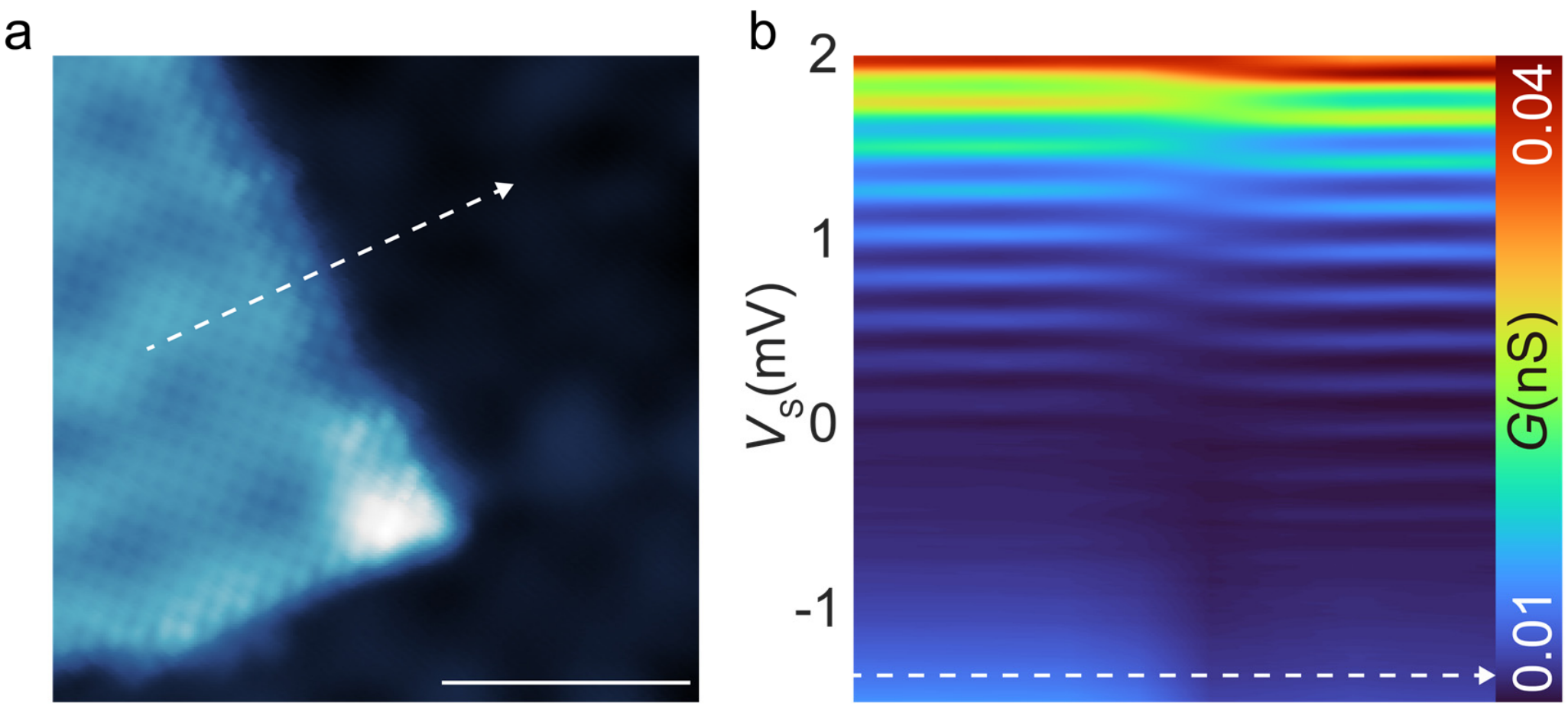

**Figure S1. $\pi$ phase shift of QWS on 2 ML NaCl.**

**a** STM image showing 2 ML NaCl on Pb underlayer. ($V_S$ = 10 mV, $I_t$ = 100 pA, scale bar is 4 nm). **b** dI/dV spectra taken along the dashed arrow in **a**. ($V_S$ = 2 V, $I_t$ = 500 pA, $V_{mod}$ = 20 mV).

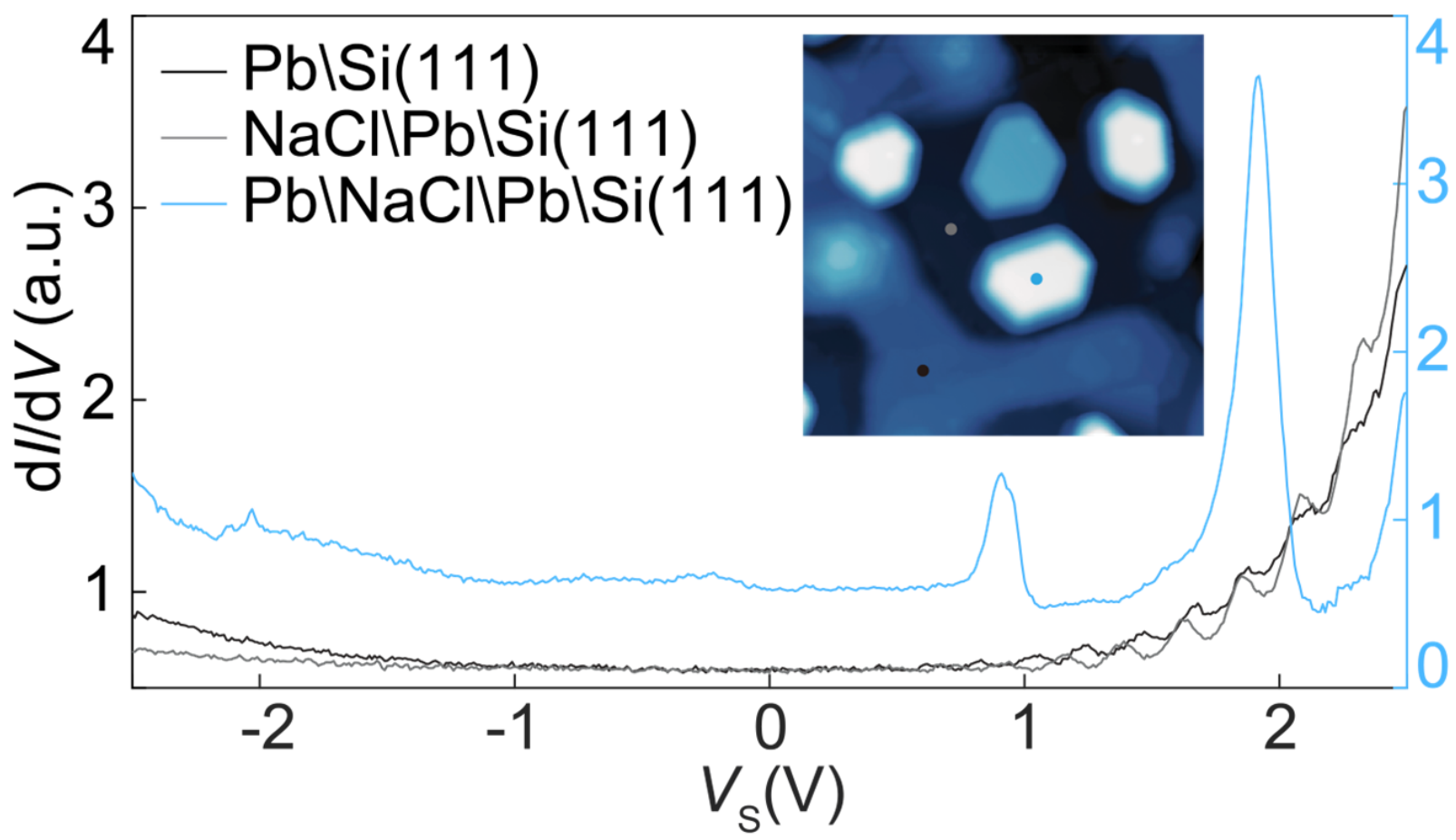

**Figure S2. Comparison of QWS between different layers.**

Point d$I$/d$V$ spectra taken on bottom, middle and top layer of the stack. ($V_S$ = 2.5 V, $I_t$ = 200 pA, $V_{mod}$ = 20 mV). Colored dots in the inset STM image indicate the positions at which the spectra were acquired.

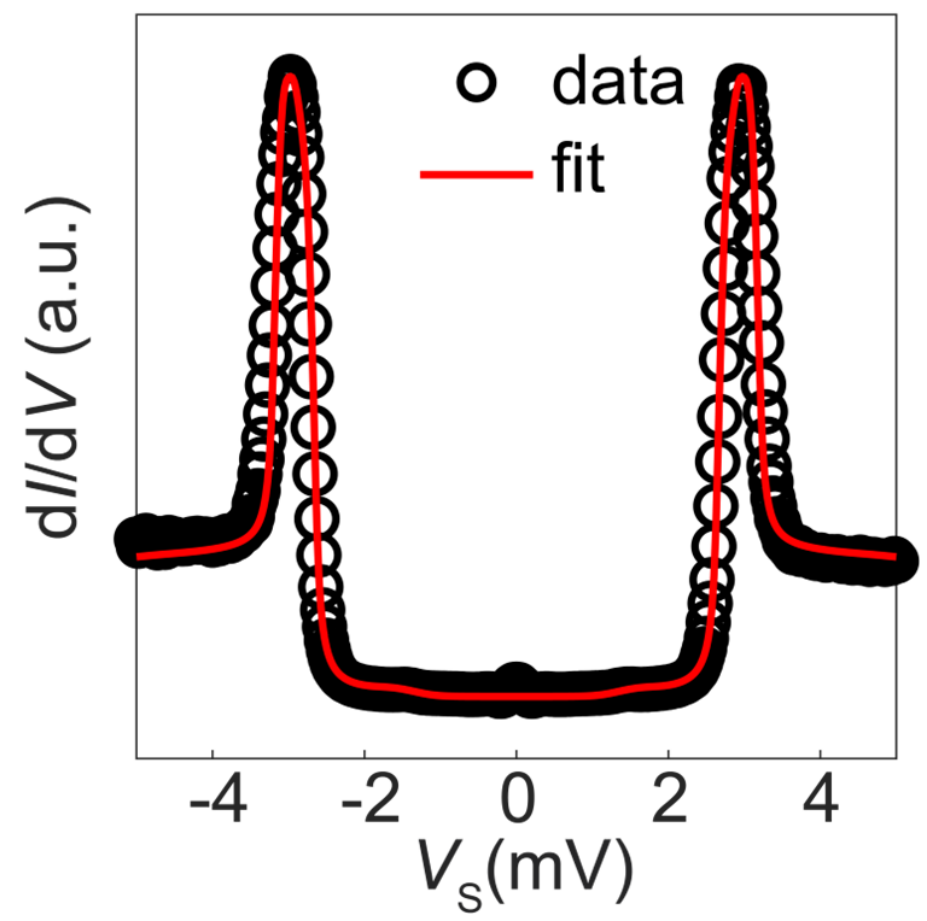

**Figure S3. Superconducting gap spectrum on Pb underlayer.**
Stabilization: $V_S$ = 5 mV, $I_t$ = 30 nA, $V_{mod}$ = 200 $\mu$V. BCS fitting parameters: $\Delta$ = 1.42 meV, $\Gamma$ = 25 $\mu$V, $T$ = 1.3 K.

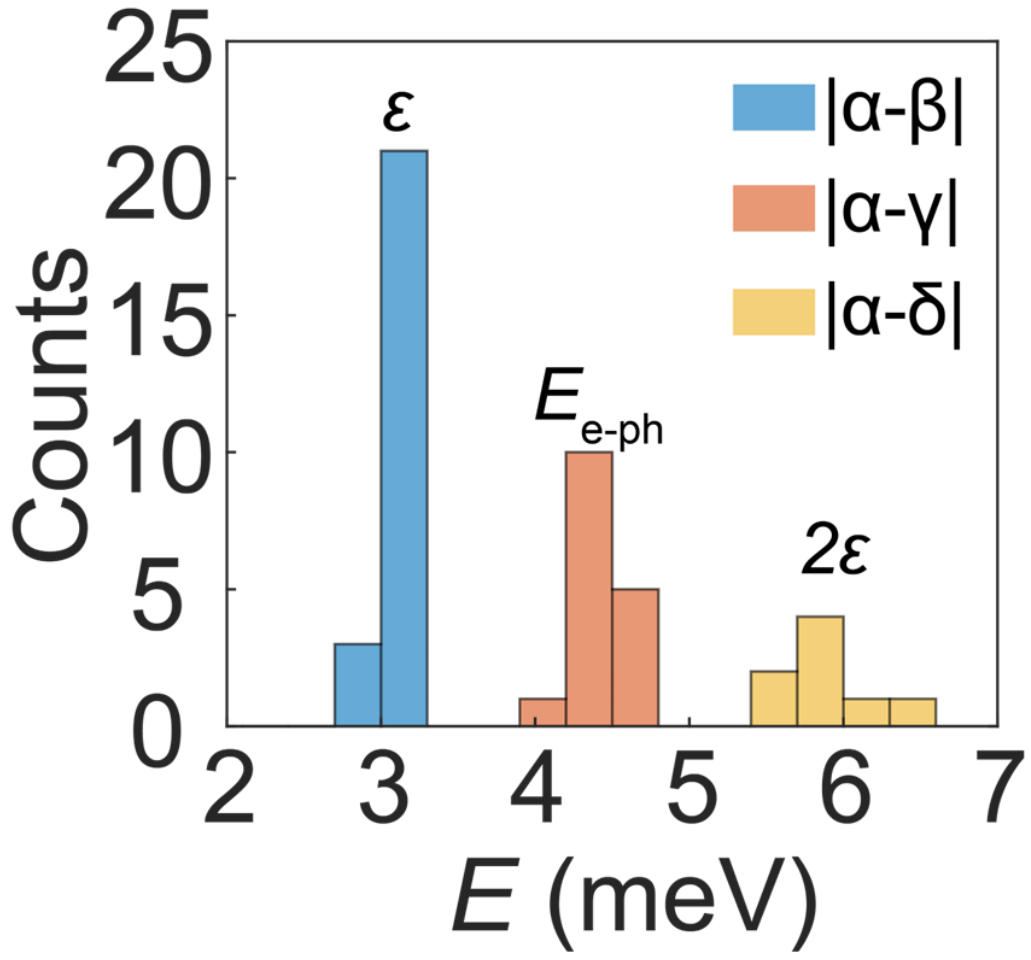

**Figure S4. Statistics on the inelastic excitation energies.**

Inelastic excitation energies were obtained from 24 stacks. The energies of all four peaks $\alpha$, $\beta$, $\gamma$ and $\delta$ were extracted by Gaussian fitting, and the excitation energies were determined from the energy differences between the inelastic peaks ($\beta$, $\gamma$ and $\delta$) and the elastic peak $\alpha$.

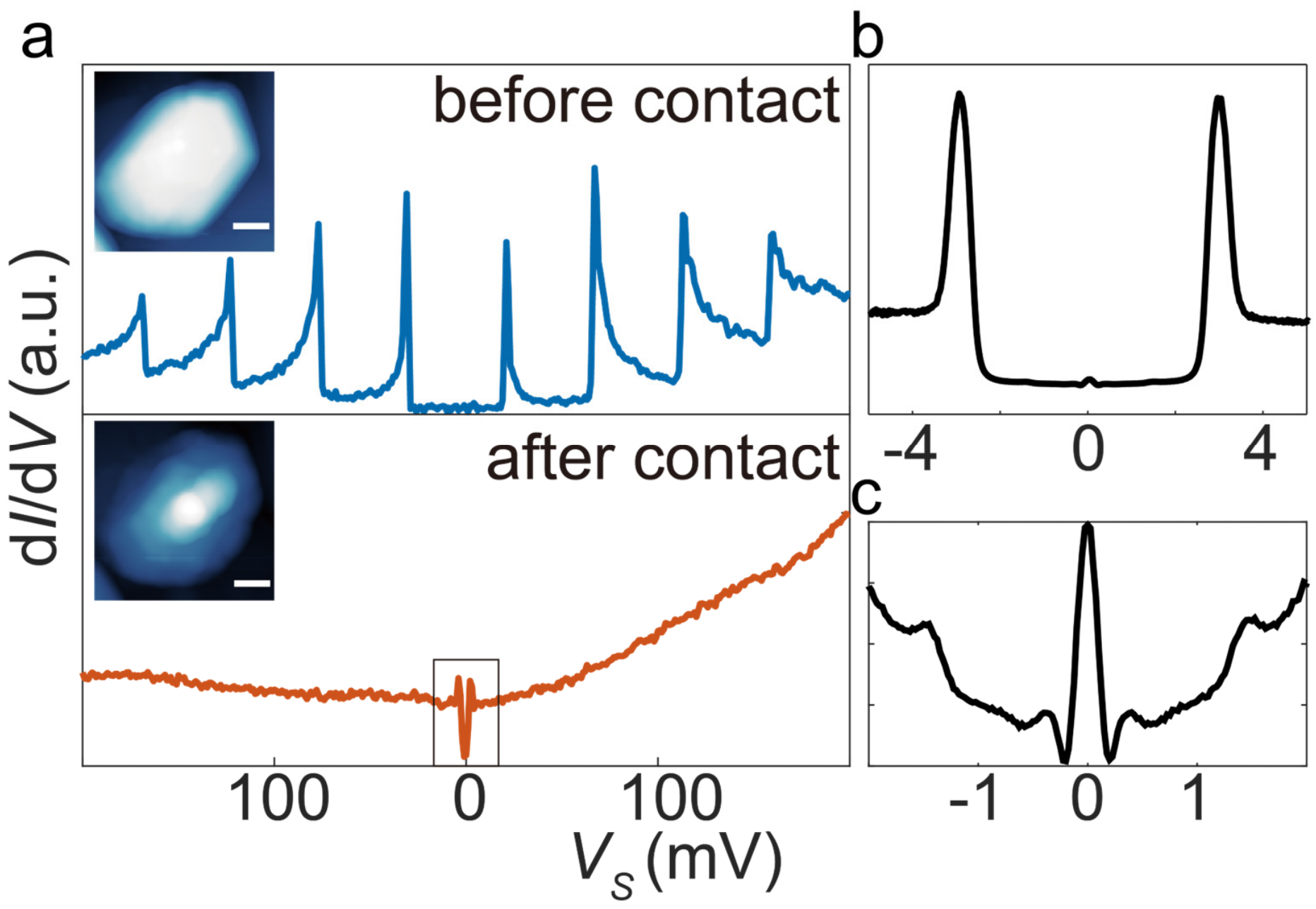

**Figure S5. Recovery of the superconducting gap in contact with the overlayer.**
**a** Upper panel: point d*I*/d*V* spectra obtained on a 10 ML Pb island before the tip contact ($V_s$ = 200 mV, $I_t$ = 200 pA, $V_{mod}$ = 1 mV). Constant-current STM image shown in the inset, scale bar = 4 nm. Lower panel: Point d*I*/d*V* spectra obtained on the same position after the tip contact ($V_s$ = 200 mV, $I_t$ = 200 pA, $V_{mod}$ = 1 mV). Constant-current STM image shown in the inset, scale bar = 4 nm. **b** Superconducting gap recovered in contact. $V_S$ = 5 mV, $I_t$ = 30 nA, $V_{mod}$ = 200 $\mu$V. **c** Zoomed-in of **b** showing the Josephson peak.

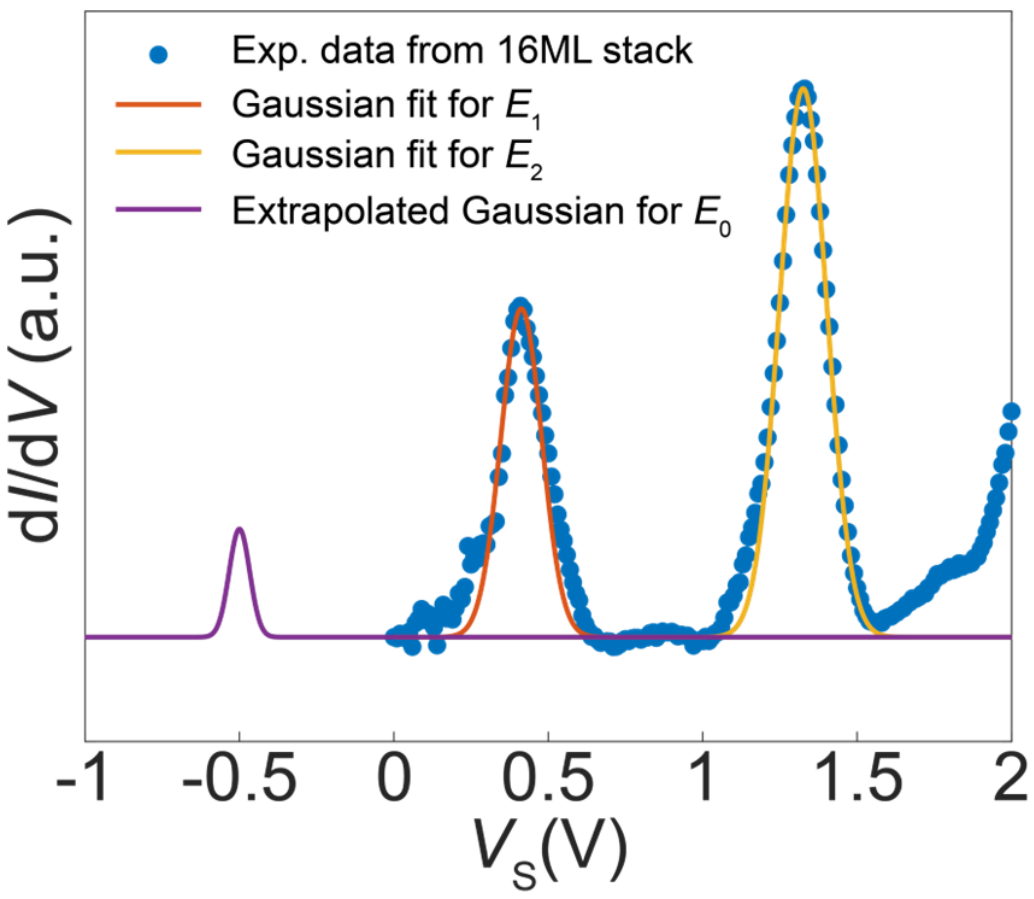

**Figure S6. Linear extrapolation of the QWS.**
Example of the linear extrapolation of the measured QWS. The energy, intensity and width for the first two QWS above $E_F$ are extracted by Gaussian fitting of the measured QWS peaks (orange and yellow curves). Following the procedure described in Supplementary Note 1, the QWS is linearly extrapolated to energies below $E_F$ (purple line).

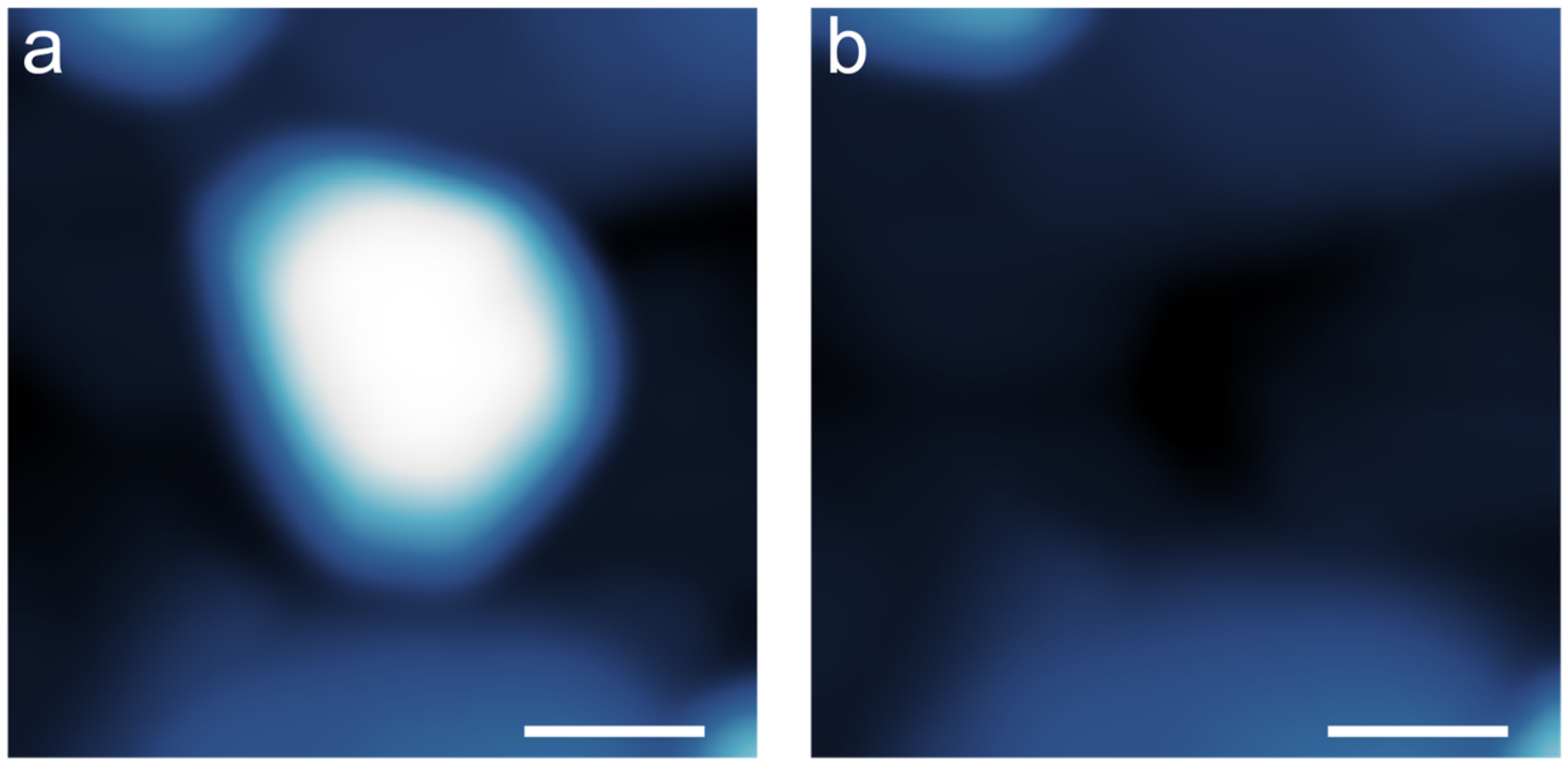

**Figure S7. Delamination of  the Pb island.**

**a** STM image of the Pb island before delamination. $V_s$ = 2.5 V, $I_t$ = 20 pA. Scale bar is 5 nm. **b** STM image in the same area after delamination. $V_s$ = 2.5 V, $I_t$ = 20 pA. Scale bar is 5 nm.